\documentclass[%
reprint,
amsmath,amssymb,
aps, nofootinbib,
floatfix, 
]{revtex4-2}

\usepackage{graphicx,physics}
\usepackage{color}
\usepackage{dcolumn}
\usepackage{latexsym}

\usepackage[normalem]{ulem}
\usepackage{hyperref,amssymb}
\usepackage{url}
\usepackage{color,soul}
\usepackage{graphicx}
\usepackage{verbatim}
\usepackage{multirow}
\usepackage{amsmath}
\newcommand{\beq}{\begin{eqnarray}}
	\newcommand{\eeq}{\end{eqnarray}}
\usepackage{mathrsfs}
\usepackage{soul}
\usepackage[dvipsnames]{xcolor}
\usepackage{mathtools}
\usepackage{slashed}
\usepackage{physics}	
\usepackage{graphicx}   
\usepackage{epstopdf}
\usepackage{subfigure}  
\usepackage{hyperref}   
\usepackage{bbold}
\usepackage{wasysym}
\usepackage{feynmp}
\usepackage{hyperref}
\usepackage{float}
\usepackage{svg}
\hypersetup{colorlinks,
}
\usepackage{bm}

\newcommand{\llangle}[1][]{\savebox{\@brx}{\(\m@th{#1\langle}\)}%
	\mathopen{\copy\@brx\mkern2mu\kern-0.9\wd\@brx\usebox{\@brx}}}
\newcommand{\rrangle}[1][]{\savebox{\@brx}{\(\m@th{#1\rangle}\)}%
	\mathclose{\copy\@brx\mkern2mu\kern-0.9\wd\@brx\usebox{\@brx}}}

\usepackage[latin1]{inputenc}

\begin{document}
	
	\title{Stress-stress correlations in two-dimensional amorphous and crystalline solids}
	\author{Jimin Bai$^{1,2,3}$}
	\author{Long-Zhou Huang$^{4}$}
	\author{Jin Shang$^{1}$}
	\author{Yun-Jiang Wang$^{4}$}
	\email{yjwang@imech.ac.cn}
	\author{Jie Zhang$^{5}$}
	\email{jiezhang2012@sjtu.edu.cn}
	\author{Matteo Baggioli$^{1,2,3}$}
	\email{b.matteo@sjtu.edu.cn}
	\address{$^1$School of Physics and Astronomy, Shanghai Jiao Tong University, Shanghai 200240, China}
	\address{$^2$Wilczek Quantum Center, School of Physics and Astronomy, Shanghai Jiao Tong University, Shanghai 200240, China}
	\address{$^3$Shanghai Research Center for Quantum Sciences, Shanghai 201315,China}
	\affiliation{
	$^{4}$State Key Laboratory of Nonlinear Mechanics, Institute of Mechanics, Chinese Academy of Sciences, Beijing 100190, China}
	\affiliation{$^{5}$Institute of Natural Sciences, Shanghai Jiao Tong University, Shanghai 200240, China}
	\begin{abstract}
		Stress-stress correlations in crystalline solids with long-range order can be straightforwardly derived using elasticity theory. In contrast, the `\textit{emergent elasticity}' of amorphous solids, rigid materials characterized by an underlying disordered structure, defies direct explanation within traditional theoretical frameworks. To address this challenge, tensor gauge theories have been recently proposed as a promising approach to describe the emergent elasticity of disordered solids and predict their stress-stress correlations. In this work, we revisit this problem in two-dimensional amorphous and crystalline solids by employing a canonical elasticity theory approach, supported by experimental and simulation data. We demonstrate that, with respect to static stress-stress correlations, the response of a 2D disordered solid is indistinguishable from that of a 2D isotropic crystalline solid and it is well predicted by \textit{vanilla} elasticity theory. Moreover, we show that the presence of pinch-point singularities in the stress response is not an exclusive feature of amorphous solids. Our results confirm previous observations about the universal character of static stress-stress correlations in crystalline and amorphous packings.
	\end{abstract}
	
	\maketitle
	\section{Introduction}
	Crystalline solids are defined by translational long-range order and their mechanical response is accurately described by elasticity theory (ET) \cite{landau2012theory,chaikin1995principles}. Theoretically, elasticity theory is based on the spontaneous breaking of translational symmetry, which implies that atoms occupy well-defined, unique equilibrium positions. These positions collectively determine the underlying crystalline structure. Stress-stress correlations in crystalline solids exhibit a characteristic power-law decay whose form is correctly captured by elasticity theory as well.
	
	Contrary to crystals, amorphous solids do not exhibit long-range translational order and their disordered structure is more similar to that of liquids, with only short-range or mid-range order \cite{binder2011glassy}. Nevertheless, amorphous solids, such as glasses, are rigid and they display several features in common with crystals, as for example long-wavelength propagating phonon modes (that nevertheless coexist with non-phononic low-energy modes \cite{10.1063/5.0069477}). Due to the absence of translational order and their inherently non-equilibrium nature, amorphous solids fundamentally challenge the core principles of elasticity theory, raising significant doubts about its applicability.
	
	With particular focus on athermal jammed solids, it has been recently proposed \cite{PhysRevLett.125.118002} (see \cite{PhysRevE.106.065004} for an extensive theoretical description) that the `\textit{emergent elasticity}' of amorphous solids can be rationalized using a particular type of tensor gauge theory \cite{PhysRevB.95.115139,PhysRevB.96.035119}, known as `\textit{vector charge theory}' (VCT) \cite{PhysRevB.100.134113}. This novel theoretical framework, inspired by the so-called fracton-elasticity dualities \cite{PhysRevLett.120.195301,PhysRevB.100.134113}, is based on the mapping of force and torque balance constraints into a generalized form of electromagnetism with a symmetric tensorial electric field $E_{ij}$. The underlying gauge symmetry emerging from this formalism reflects the lack of a unique stress-free equilibrium configuration in amorphous solids that renders the traditional concepts of displacement vector and strain tensor ill-defined. In the language of VCT, the displacement vector becomes an (unphysical) potential $\phi_i$ for the tensorial electric field that is, on the other hand, dual to the measurable and physical stresses $\sigma_{ij}$ in these amorphous solids. By considering a VCT in a dielectric, the stress-stress correlations in zero temperature amorphous solids have been theoretical predicted and successfully matched to experimental and simulation data in 2D and 3D \cite{PhysRevLett.125.118002,PhysRevE.106.065004,10.1063/5.0131473,10.3389/fphy.2022.1048683,countryman2025pinchpointsingularitiesstressstresscorrelations}.
	
	Despite the success of VCT in describing the emergent (static) elasticity of amorphous solids, several open questions remain. In particular, the similarities with elasticity theory are in fact striking, in particular when ET is considered in its stress-only formulation. Is VCT really needed to derive the static stress-stress correlations in amorphous solids? Which would be the predictions of \textit{vanilla} elasticity theory \cite{landau2012theory,chaikin1995principles} and how would they differ from those arising from VCT?

	In considering these questions, we notice that (I) the idea of getting rid of the displacement vector and the strain tensor is already imprinted in the stress-only formulations of elasticity theory, that are ultimately equivalent to canonical ET. (II) Despite a free energy functional is never invoked in VCT, a Ginzburg-Landau action, with the exact same quadratic form, is used. (II) VCT cannot be completed and be predictive without a direct match to elasticity theory and without identifying the dieletric tensor (entering in the Ginzburg-Landau functional) with the inverse elastic tensor.
	
	Apart from these theoretical aspects, we notice that stress-stress correlations in amorphous solids had been already derived \cite{PhysRevE.96.052101,10.1063/1.5041461} (see also \cite{10.1063/5.0034728}) by only imposing material isotropy and mechanical balance, without any specifications of the material microscopic properties. This suggests that the same formulae should also apply to isotropic crystals (\textit{e.g.}, 2D crystals with triangular lattice). In fact, the striking universality of static stress-stress correlations in the long-wavelength limit in near-crystalline and amorphous solids has been recently observed using simulations in \cite{PhysRevE.109.044903}.

	The purpose of this work is to revisit the question of static stress-stress correlations in 2D amorphous and crystalline systems. As we will explicitly demonstrate and corroborate with experimental and simulation data, the VCT predictions for the static stress-stress correlations are identical to those of \textit{vanilla} elasticity theory for isotropic crystals \cite{landau2012theory,chaikin1995principles}. In other words, the static stress response of amorphous solids appears of the same form of that of isotropic crystals, including the presence of pinch-point singularities that were previously ascribed as exclusive characteristic of amorphous solids (see, e.g., \cite{PhysRevE.106.065004} and references therein).
	
	We anticipate that this surprising universality pertains only the static stress-stress correlations and does not extend to other observables and in particular to the dynamical response. We also clarify that all our discussion is based on the long-wavelength limit in which elasticity theory is valid and applies only in the regime where elastic screening is negligible (\textit{e.g.}, large pressure regime in athermal granular systems). Away from such regime, differences between amorphous systems and crystalline material obviously emerge. That is not of interest for the present discussion. 
	
	\section{Stress-stress correlations from vanilla elasticity theory}\label{thth}
    Elasticity theory \cite{chaikin1995principles,landau2012theory} is a continuous framework built upon a symmetric rank $2$ tensor $u_{ij}$ (\textit{strain tensor}) that encodes the mechanical deformations of a material. In the linear approximation, and in flat space, this tensor can be written as:
    \begin{equation}\label{nono}
u_{ij}=\partial_{(i}d_{j)}=\frac{1}{2}\left(\partial_i d_j+\partial_j d_i\right)
    \end{equation}
    that is that symmetrized derivative of the displacement vector $d_i$. Expressing $u_{ij}$ in terms of the displacement $d_i$ is possible only when the Saint-Venant's compatibility condition is satisfied.

Importantly, it follows from the constitutive relation that the stress tensor $\sigma_{ij}$ is linearly proportional to the strain tensor
\begin{equation}
		\sigma_{ij} = \mathcal{E}_{ijkl} u_{kl},
		\label{eq:constitution1}
	\end{equation}
    where $\mathcal{E}_{ijkl}$ is a rank $4$ tensor that contains the information about the elastic constants of the material, hence, it is named the \textit{elastic tensor}. Eq. \eqref{eq:constitution1} is simply a formal rewriting of Hooke's law (\textit{ut tensio, sic vis}), namely that the elastic force is proportional to the displacement. As we will see in detail below, the precise structure of this tensor depends on the symmetries of the crystalline lattice. Nevertheless, $\mathcal{E}_{ijkl}$ is always\footnote{This is not anymore true in systems exhibiting odd elasticity or local rotational degrees of freedom (micropolar elasticity).} symmetric under the exchange of the first/last  two indices and also under interchange of the first and second pairs of indices. These symmetries reduce the number of independent entries in $\mathcal{E}_{ijkl}$. 

The elastic free energy $F_{\text{el}}$ can be then written at leading order as a quadratic form in the strain,
\begin{equation}
    F_{\text{el}}= \frac{1}{2}\int d^dx \,u_{ij} \mathcal{E}_{ijkl} u_{kl}=\int d^dx f_{\text{el}}\left[u_{ij} \right].\label{free}
\end{equation}
The functional derivative of the free energy density $f_{\text{el}}$ with respect to the strain (at constant temperature) defines the conjugate stress,
\begin{equation}
    \sigma_{ij}\equiv \frac{\delta f_{\text{el}}\left[u_{ij} \right]}{\delta u_{ij}},
\end{equation}
recovering the constitutive equation, Eq. \eqref{eq:constitution1}. This also implies that the free energy density can be equivalently written as
\begin{equation}
    f_{\text{el}}=  \frac{1}{2}\sigma_{ij} u_{ij}=  \frac{1}{2} \sigma_{ij} \left(\mathcal{E}^{-1}\right)_{ijkl} \sigma_{kl},
\end{equation}
where $\mathcal{E}^{-1}$ is the inverse of the elastic tensor.

In absence of external forces, \textit{i.e.} at equilibrium, the stress tensor satisfies
	\begin{equation}
		\partial_i \sigma_{ij} = 0.
		\label{eq:force free}
	\end{equation}
In two dimensions, Eq. \ref{eq:force free} is automatically satisfied after defining the Airy potential $\psi$,
	\begin{equation}
		\sigma_{ij}=
			\epsilon_{ia}\epsilon_{jb} \partial_{a} \partial_{b} \psi,
		\label{eq:potential_form}
	\end{equation}
    where $\epsilon_{ij}$ is the anti-symmetric Levi-Civita tensor in 2D.
	In wave-vector space (q-space, $\{q_x,q_y\}$), Eq. \ref{eq:potential_form} becomes
	\begin{equation}
		\sigma_{ij}(q) = \epsilon_{ik}\epsilon_{jl}q_kq_l\psi(q).
		\label{eq:Airy}
	\end{equation}
    At this point, it is convenient to use Voigt notation by denoting $\tilde{\mathcal{E}}_{ij} \equiv \mathcal{E}_{ijkl}$, $\tilde{\sigma}_i \equiv \sigma_{ij}$, following the convention that the indices in the tilde quantities follow the order \{xx, yy, xy\}. Using Voight notation, Eq. \ref{eq:Airy} can be expressed as
	\begin{equation}
		\tilde{\sigma}_i(q) = A_i(q)\psi(q),
	\end{equation}
	where, in two dimensions,
	\begin{equation}
		A_i(q) = (q_y^2, q_x^2, -q_xq_y).
	\end{equation}
    From the elastic free energy, we can define the effective Gaussian action $S$ that in Voigt notation takes the following form:
    \begin{equation}
S[\psi] = \frac{1}{2}\int d^2q \,\left(\tilde{\mathcal{E}}^{-1}\right)_{ij}A_i(q)A_j(-q)\psi(q)\psi(-q).
\label{eq:Hamilt}
    \end{equation}
The correlation function of $\psi(q)$ can be directly obtained by functional derivation
\begin{equation}
   	\begin{aligned}
   		\langle\psi(q)\psi(-q)\rangle &= \frac{\int [\mathcal{D}\psi]e^{-S/g}\psi(q)\psi(-q)}{Z}\\
   		&= \left[A_i(q)\left(\tilde{\mathcal{E}}^{-1}\right)_{ij}A_j(-q)\right]^{-1}.
   		\label{eq:correlator_core}
   	\end{aligned}
   \end{equation}

The ensemble average in Eq. \eqref{eq:correlator_core} refers to the thermodynamic ensemble that is defined by the Boltzmann probability distribution $e^{-\frac{S}{k_BT}}$. In other words, in thermal systems, $g \equiv k_B T$. Since the value of $g$ affects our final results only up to an overall constant factor, for simplicity, we will set $g=1$ in the rest of the manuscript. Before proceeding, we notice that the choice of the ensemble over which the average is performed is the main difference between the computation from elasticity theory (sketched in this section) and the analogous computation for athermal granular systems presented in \cite{PhysRevE.106.065004}. There, the ensemble average is not on the Boltzmann thermal distribution and $g \neq k_B T$. In that case, the average is performed using an equiprobability ansatz that is the basis of the Edwards stress ensemble (see \cite{annurev:/content/journals/10.1146/annurev-conmatphys-031214-014336} for a recent review). The Edwards ensemble is based on the fact that volume is conserved and compactivity $\chi_V$ equilibrates, as temperature does in the Boltzmann ensemble. One can then compute the density of jammed states $\Omega(V)$ as a function of the volume $V$ and, finally, build the Edwards entropy, the conjugate variable of the compactivity $\chi_V$. This framework is particularly useful for jammed granular matter \cite{RevModPhys.90.015006}, and it has been recently experimentally validated in three-dimensional granular systems \cite{PhysRevLett.127.018002}. As we will concretely see, this difference will not alter the form of the correlations as a function of the wave-vector $\textbf{q}$ nor the angle $\theta$.

Using Eq. \eqref{eq:correlator_core}, the correlations of the stress fields $\tilde{\sigma}$ finally can be written
   \begin{equation}
   	\begin{aligned}
   		\langle\tilde{\sigma}_i\tilde{\sigma}_j\rangle &= A_i(q)A_j(q)\langle\psi(q)\psi(-q)\rangle\\
   		&= A_i(q)A_j(q)\left[A_k(q)\left(\tilde{\mathcal{E}}^{-1}\right)_{kl}A_l(-q)\right]^{-1}.
   	\end{aligned}
   	\label{eq:ss_corre}
   \end{equation}
Up to an overall constant factor, Eq. \eqref{eq:ss_corre} gives the prediction for the stress correlations in classical elasticity theory that can be directly computed once the form of the elastic tensor $\mathcal{E}$ is specified. We notice that, in order to derive the results, we have made no use of Eq. \eqref{nono}. In the rest of this work, we are interested in the correlations of the stress fluctuations
   \begin{equation}
   	\delta\sigma_{ij} = \sigma_{ij}-\langle\sigma_{ij}\rangle
   \end{equation}
around the average value $\langle\sigma_{ij}\rangle$. We will also indicate the correlation functions of the stress fluctuations as
\begin{equation}
    C_{ijkl}\equiv \langle \delta \sigma_{ij} \delta \sigma_{kl}\rangle.
\end{equation}
  \subsection{Stress-Stress Correlations for a 2D Triangular Lattice}
  A 2D triangular lattice possesses 6-fold symmetry. This reduces the number of independent non-zero elasticity constants to 2. The elastic free energy density can then expressed as
  \begin{equation}
  	f_{\text{el}}= \frac{K_A}{2}(u_{xx}+u_{yy})^2+\frac{\mu}{2}\left[(u_{xx}-u_{yy})^2+4u_{xy}^2\right],
  	\label{free energy 6 fold 2D}
  \end{equation}
  with the area compression (bulk) modulus $K_A$ and the shear modulus $\mu$. We notice that this form is identical to that of an isotropic 2D solid. In Voigt notation, the elasticity tensor for a 2D triangular lattice is then
  \begin{equation}
  	\tilde{\mathcal{E}}=\begin{bmatrix}
  		K_A + \mu & K_A - \mu & 0 \\
  		K_A - \mu & K_A + \mu & 0 \\
  		0 & 0 & 2\mu
  	\end{bmatrix}.
  	\label{eq:IsotropicElasticity}
  \end{equation} 
It follows from Eq. \ref{eq:ss_corre} that the stress-stress correlations are given by the simple expression
  \begin{equation}
  	\langle\psi(q)\psi(-q)\rangle = 4K_{2D}~(q_x^2+q_y^2)^{-2} = 4K_{2D} |q|^{-4},
  	\label{correlator}
  \end{equation}
  with
\begin{equation}
  	K_{2D} =\frac{K_A+\mu}{4K_A\mu}.
  \end{equation}
  By denoting $sin(\theta)=q_y/q$, $cos(\theta)=q_x/q$, we find the final form:
  \begin{equation}
  	\begin{aligned}
  		&\langle \delta\sigma(q)_{i} \delta\sigma(-q)_{j} \rangle =  K_{\text{2D}} * \\& \begin{bmatrix} 4 \sin ^4(\theta ) &  \sin ^2(2 \theta ) & -4  \sin ^3(\theta ) \cos (\theta ) \\
  			 \sin ^2(2 \theta ) & 4  \cos ^4(\theta ) & -4 \sin (\theta ) \cos ^3(\theta ) \\
  			-4 \sin ^3(\theta ) \cos (\theta ) & -4 \sin (\theta ) \cos ^3(\theta ) &\sin ^2(2
  			\theta )
  		\end{bmatrix}.\\
  	\end{aligned}
  	\label{eq:isotropicCorrel}
  \end{equation}
  This equation is identical to that predicted from tensor gauge theories and utilized to match the stress correlations in 2D amorphous solids (see \textit{e.g.} Eq. (A27) in \cite{PhysRevE.106.065004}). As evident from our derivation, pure elasticity theory predicts the same form for an isotropic 2D crystal. The only difference between the two frameworks is the choice of the configuration ensemble, thermal in one case and athermal in the other. This difference does not alter the form of the correlations but only affects the overall constant pre-factor.
  \subsection{Stress-Stress Correlations for a 2D Square Lattice}
  A crystal with a 2D square lattice structure possesses 4-fold symmetry. This requires  $\mathcal{E}_{xxxx}=\mathcal{E}_{yyyy}$, leaving the number of independent non-zero elasticity constants to 3.
  
  The general form of the elastic free energy density for a square lattice is:
  \begin{equation}
   f_{\text{el}} = \frac{K_A}{2}(u_{xx}+u_{yy})^2+\frac{\mu_p}{2}(u_{xx}-u_{yy})^2+2\mu_s u_{xy}^2,
  	\label{free energy 6 fold 2D}
  \end{equation}
  with the area compression modulus $K_A$ and two shear moduli $\mu_p$ and $\mu_s$.
  
  In Voigt notation, the elasticity tensor in this case becomes:
\begin{equation}
  	\tilde{\mathcal{E}}=\begin{bmatrix}
  		K_A + \mu_p & K_A - \mu_p & 0 \\
  		K_A - \mu_p & K_A + \mu_p & 0 \\
  		0 & 0 & 2\mu_s
  	\end{bmatrix}.
  \end{equation} 
 Similarly, for the stress-stress correlations, we find from Eq. \ref{eq:ss_corre},
\begin{equation}
  	\begin{aligned}
  		&\langle\psi(q)\psi(-q)\rangle^{-1} \\
  		&=  \frac{(K_A+\mu_p)}{4K_A\mu_p}\left(q_x^2+q_y^2\right)^2  +  \left(\frac{1}{\mu_s}-\frac{1}{\mu_p}\right)q_x^2q_y^2.
  	\end{aligned}
  	\label{correlator : square lattice}
  \end{equation}
By denoting $sin(\theta)=q_y/q$, $cos(\theta)=q_x/q$, $tan(\theta)=q_y/q_x$, we obtain the final form:
  \begin{widetext}
  	\begin{equation}
  		\langle \delta\sigma(q)_{i} \delta\sigma(-q)_{j} \rangle =
  		\begin{bmatrix}  
  			\frac{1}{C_A\sin^{-4}\theta + C_B\tan^{-2}\theta}  &  \frac{1}{C_A\sin^{-2}\theta \cos^{-2}\theta + C_B} & \frac{-1}{C_A\sin^{-3}\theta \cos^{-1}\theta + C_B\tan^{-1}\theta} \\
  			\frac{1}{C_A\sin^{-2}\theta \cos^{-2}\theta + C_B} & \frac{1}{C_A\cos^{-4}\theta + C_B\tan^{2}\theta} & \frac{-1}{C_A\sin^{-1}\theta \cos^{-3}\theta + C_B\tan\theta} \\
  			\frac{-1}{C_A\sin^{-3}\theta \cos^{-1}\theta + C_B\tan^{-1}\theta} & \frac{-1}{C_A\sin^{-1}\theta \cos^{-3}\theta + C_B\tan\theta} & \frac{1}{C_A\sin^{-2}\theta \cos^{-2}\theta + C_B}
  		\end{bmatrix},
  		\label{eq:SquareCorrel}
  	\end{equation}
  \end{widetext}
  with the coefficients $C_A$ and $C_B$ defined through
  \begin{equation}
  		C_A = \frac{(K_A+\mu_p)}{4K_A\mu_p},\qquad
  		C_B = \frac{1}{\mu_s}-\frac{1}{\mu_p}.
  	\label{eq:SquareCoeff}
  \end{equation}

Before continuing, we clarify that our work does not introduce new theoretical models but rather relies on established results from standard elasticity theory. For athermal amorphous solids, more sophisticated theories, able to derive also an ``equation of state'', have been already developed \cite{PhysRevLett.121.118001,PhysRevE.98.033001}. Numerical studies of stress-stress correlations in deeply supercooled liquids also exist, e.g., \cite{PhysRevLett.113.245702}.
    \begin{figure}[htbp]
			\centering
			\includegraphics[width=1.0\linewidth]{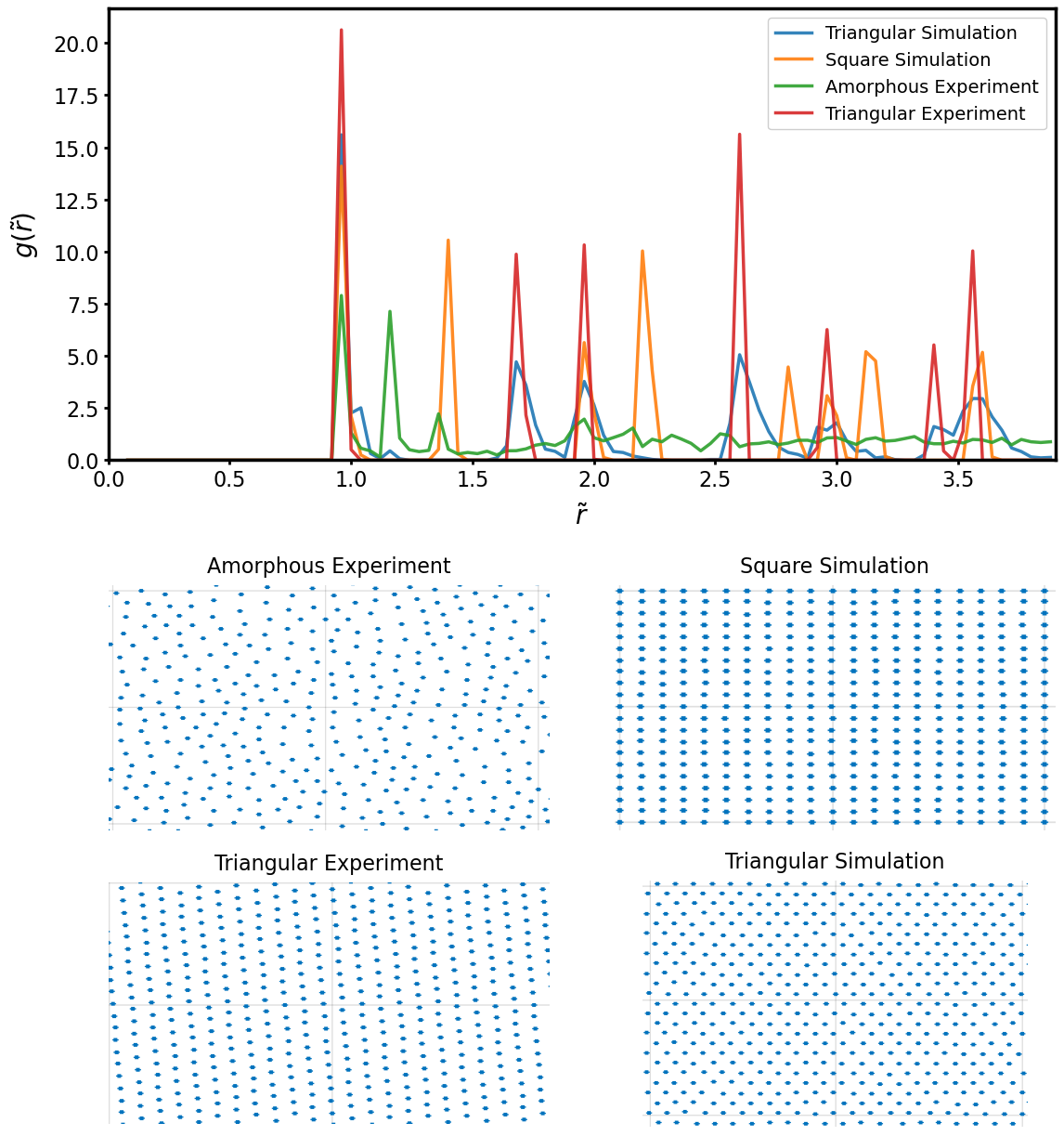}
			\caption{Pair distribution function $g(\tilde r)$: (TLS) triangular simulation, (SLS) square simulation, (AE) amorphous experiment, and (TLE) triangular experiment. $\tilde r$ denotes the radial length-scale normalized by the nearest neighbor distance for crystalline systems and the average distance between nearby particles for the amorphous packing. The images below provide snapshots of the corresponding structures.}
			\label{fig:gr}
		\end{figure}
	\section{Experimental system and simulations}
	\subsection{Experimental details}\label{exp}
    
\begin{figure}[htbp]
		\centering
		\includegraphics[width=0.8
    \linewidth]{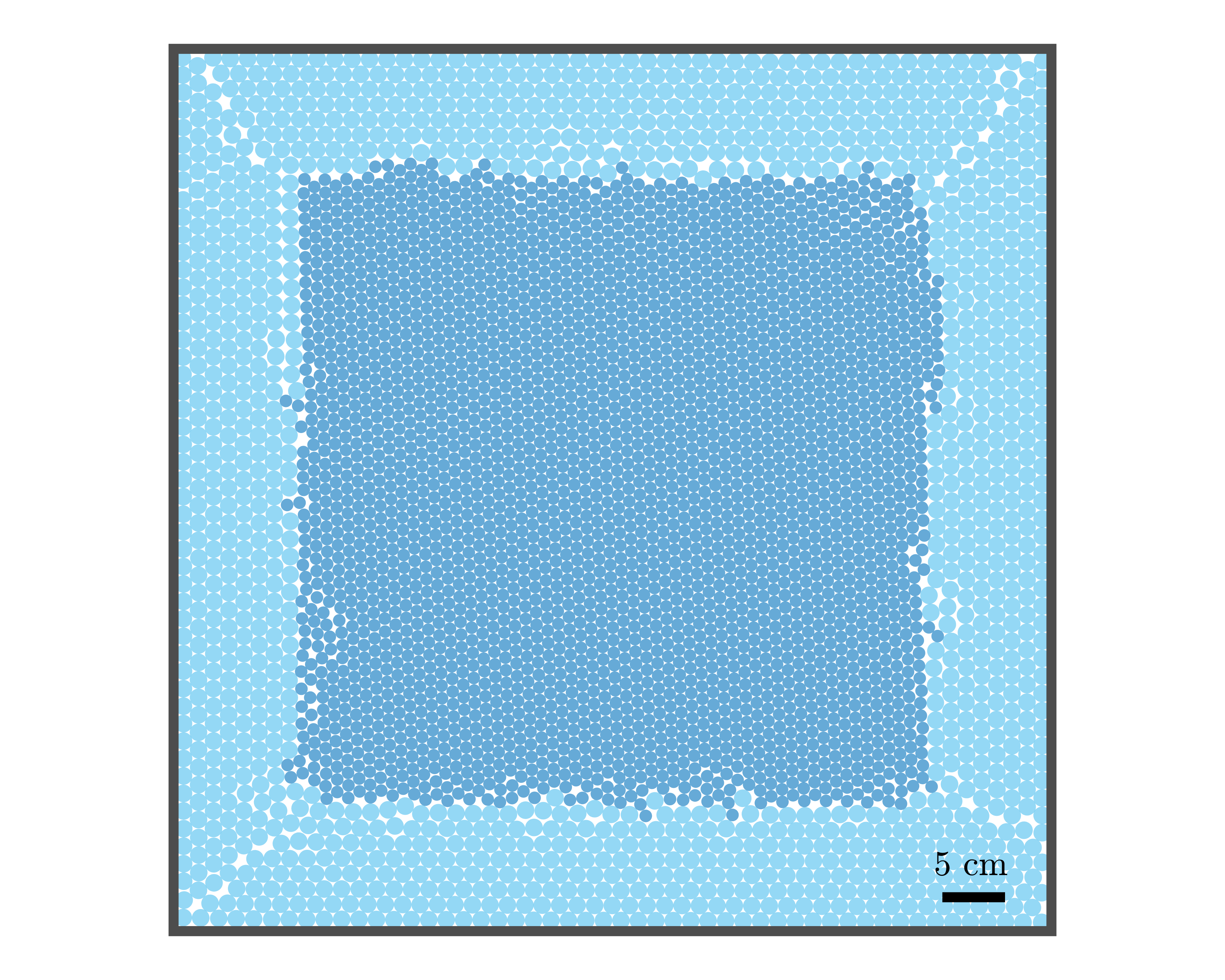}
		\caption{A typical example of the experimental particle configuration. The central dark blue circles represent small disks with 10 mm diameter, forming a triangular lattice. The surrounding light blue circles denote the large disks with 14 mm diameter. Irregular grain boundaries exist between small and large disks.}
		\label{fig:exset}
	\end{figure}

    The 2D granular experiment is conducted on a horizontal glass plate, with a movable square boundary of an initial length of 693 mm. For the crystal case, 2880 small disks (10 mm diameter) are arranged in a triangular lattice from the center of the system. In order to minimize the influence of boundary flatness on crystallization, the crystal of small disks is surrounded by 8 layers of larger disks (14 mm diameter). The large disks contact the boundary, while the small central particles formed irregular grain boundaries with the large particles, as shown in Fig.~\ref{fig:exset}. Due to the random structure of the grain boundaries, the stress within the internal crystal exhibits space fluctuations. Our analysis only focus on well-crystallized regions of the internal crystal.
    
    In the experiments, we apply photoelastic techniques to measure the inter-particle forces. Initially, we adjust the structure arrangement of the crystals and grain boundaries to achieve zero pressure. Subsequently, isotropic quasi-static compression is applied via boundary displacement, with a maximum compressive ratio of 1.04\%. Following this protocol, we obtain 7 independent samples. For further information of the experimental apparatus and the photoelastic techniques, please refer to Refs. \cite{wangyinqiao2020NC,wangyinqiao2021PRR}.

    For the amorphous case, the bidisperse packing consists of 2710 small and 1355 large disks. In the initial state, the randomly mixed disks are placed within a square boundary of 711 mm length, with a packing fraction near the jamming point. Then we compress the system in the same manner, with a maximum compress ratio of 1.69\%.

    \subsection{Simulation details}\label{sim}
    Two dimensional simulation systems with 22,400 particles for the triangular lattice and 10,000 particles for the square lattice are constructed by using periodic boundary conditions. Reduced units are employed. The simulation time step is 0.001. The Johnson-Kendall-Roberts model is used to mimic the normal contact \cite{doi:10.1098/rspa.1971.0141}. The normal elastic component of force acting on the particle $i$ due to contact with particle $j$ is given by 
    \begin{equation}
        \mathbf{F}_{\mathrm{ne,jkr}}=\left(\frac{4 E_{\mathrm{eff}} a^{3}}{3 R}-2 \pi a^{2} \sqrt{\frac{4 \gamma E_\mathrm{eff}}{\pi a}}\right) \mathbf{n},
    \end{equation}
    where
    \begin{equation}
        E_{\mathrm{eff}}=\left(\frac{1-\nu_{i}^{2}}{E_{i}}+\frac{1-\nu_{j}^{2}}{E_{j}}\right)^{-1}
    \end{equation}
is the effective Young's modulus. $E$ takes the value $10^{5}$ in the triangular lattice and $10^{6}$ in the square lattice. $\nu=0.3$ is the Poisson ratio; $a$ is the radius of the contact zone, $\gamma=50$ is the surface energy density; and $\mathbf{n}=\frac{\mathbf{r}_{ij}}{\|\mathbf{r}_{ij}\|}$, where $\mathbf{r}_{ij}=\mathbf{r}_{i}-\mathbf{r}_{j}$ is the position vector of the two particles. 

In addition, the normal force is enhanced by viscoelastic damping terms \cite{PhysRevE.53.5382}:
\begin{equation}
    \mathbf{F}_{\mathrm{n,damp}}=-\eta_{\mathrm{n}}am_{\mathrm{eff}}\mathbf{v}_{\mathrm{n,rel}},
\end{equation}
where $\eta_{\mathrm{n}}=0.1$ is a damping prefactor and $m_{\mathrm{eff}}=m_im_j/(m_i+m_j)$ is the effective mass. Finally, $\mathbf{v}_{\mathrm{n,rel}}$ is relative velocity. All in all, the total normal force is computed as the sum of the elastic and damping components: $\mathbf{F}_\mathrm{n}=\mathbf{F}_{\mathrm{ne}}+\mathbf{F}_{\mathrm{n,damp}}$. Tangential Mindlin model is used to tangential contact \cite{Mindlin1949ComplianceOE}, the tangential force is given by
\begin{equation}
    \mathbf{F}_\mathrm{t}=-\min(\eta_{\mathrm{t}}\|\mathbf{F}_\mathrm{n}\|,\|-8G_{\mathrm{eff}}\xi+\mathbf{F}_{\mathrm{t,damp}}\|)\mathbf{t},
\end{equation} 
where
\begin{equation}
    G_{\mathrm{eff}}=\left(\frac{2-\nu_i}{G_i}+\frac{2-\nu_j}{G_j}\right)^{-1}
\end{equation}
is the effective shear modulus and $G_i=E_i/(2(1+\nu_i))$ is the shear modulus. Also, $\xi$ is the tangential displacement accumulated during the entire duration of the contact, $\eta_{\mathrm{t}}=0.5$ is the tangential friction coefficient and $\mathbf{F}_{\mathrm{t,damp}}$ is the tangential damping force. Rolling friction is computed via a spring-dashpot-slider \cite{Luding}, the rolling pseudo-force is given by $\mathbf{F}_\mathrm{roll,0}=k_\mathrm{roll}\xi_\mathrm{roll}-\gamma_\mathrm{roll}\mathbf{v}_\mathrm{roll}$. Here, $k_\mathrm{roll}=10^3$ is rolling stiffness, $\xi_\mathrm{roll}$ is the rolling displacement, $\gamma_\mathrm{roll}=10^3$ is the rolling damping and $\mathbf{v}_\mathrm{roll}$ is the relative rolling velocity; the Coulomb friction criterion truncates the rolling pseudo-force when it surpasses a critical threshold: $\mathbf{F}_\mathrm{roll}=\min(\mu_\mathrm{roll}\|\mathbf{F}_\mathrm{n}\|,\|\mathbf{F}_\mathrm{roll,0}\|)\mathbf{k}$, where $\mu_\mathrm{roll}=0.1$ is the rolling friction coefficient and $\mathbf{k}=\mathbf{v}_\mathrm{roll}/\|\mathbf{v}_\mathrm{roll}\|$ is the direction of the pseudo-force. 

We refer to the LAMMPS documentation for more details \cite{lammps}. The kinetic properties were estimated using standard microcanonical (NVE) ensemble after running 10,000 time steps under temperature 0.001. For statistical purpose, twenty independent samples were simulated to reduce the fluctuation of physical quantities. The velocity Verlet algorithm  was used to integrate Newton's equations of motion. LAMMPS were used for molecular dynamics simulation and OVITO \cite{ovito} was used for visualization of the model. We also remind that the temperature in the simulations is set to a low value $T\approx 0.001$ in reduced units. 
\begin{figure}[htbp]
\includegraphics[width=1.0\linewidth]{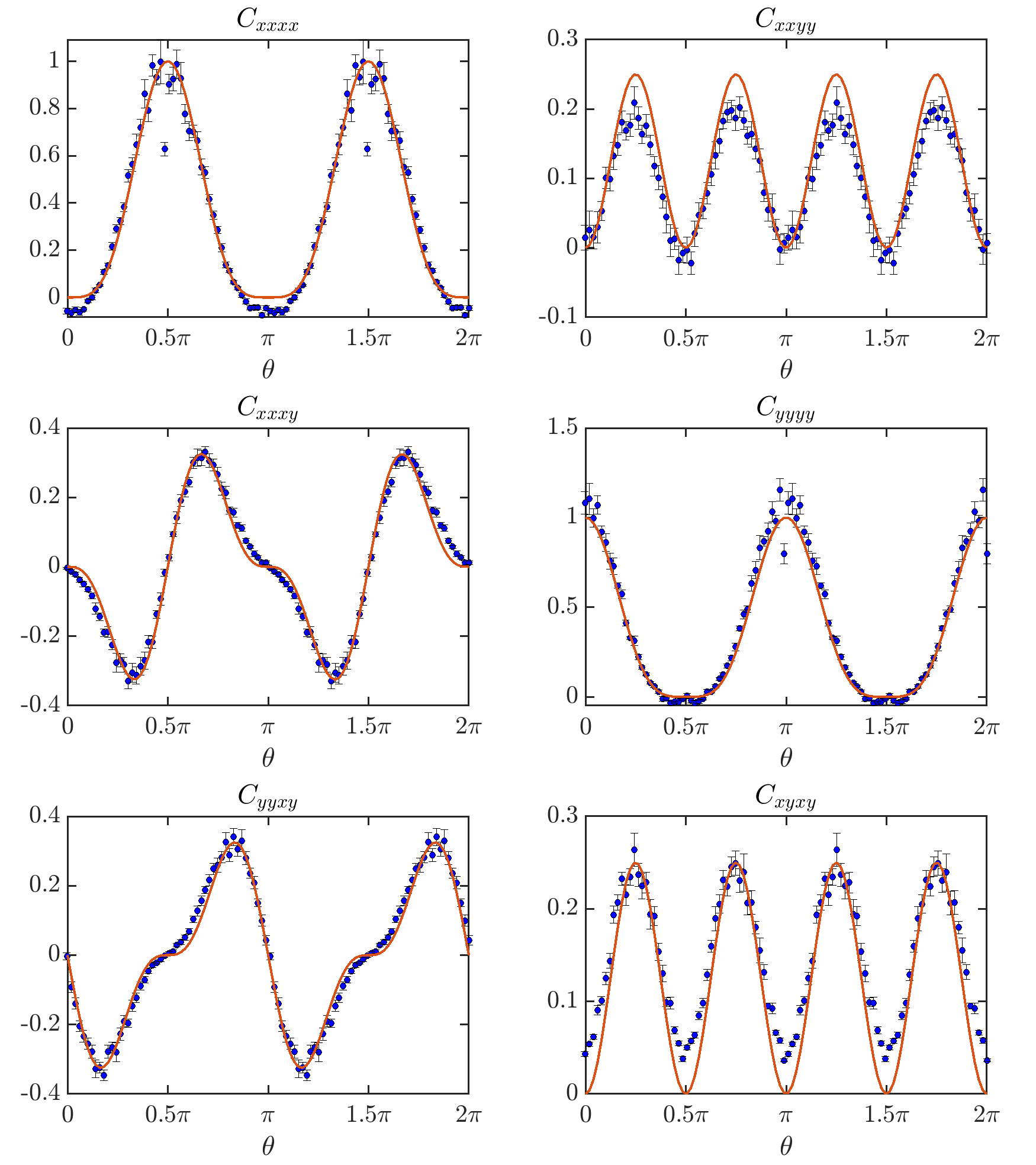}
\caption{Normalized stress fluctuation correlations $C_{ijkl}(\theta)$ obtained from the amorphous bidisperse experimental packing (AE in Fig. \ref{fig:gr}). The data are obtained by integrating the stress fluctuation correlations $C_{ijkl}$ over $q$ to retain only the angular dependence, where $q$ is the magnitude of the Fourier space $q$-vector. The results are the ensemble average over $10$ independent experimental datasets and the error bars indicate the corresponding statistical errors. The solids lines are the theoretical predictions from Eq. \eqref{eq:isotropicCorrel}.}
    \label{fig:1DAE}
\end{figure}

\section{Amorphous packing versus isotropic crystal}
Our theoretical analysis suggests that stress correlations in 2D amorphous packings and in isotropic 2D crystals should take the same form as a function of the angle $\theta$. In order to corroborate this, we resort to the experimental and simulation methods described above.

We start by considering a 2D experimental amorphous packing (see details in Section \ref{exp}). A snapshot of the amorphous structure is presented in the inset of Fig. \ref{fig:gr} (AE). To confirm the amorphous nature, we computed the pair distribution function $g(r)$ in units of the average distance between nearby particles. The result in Fig. \ref{fig:gr} confirms the absence of long-range order.

In Fig. \ref{fig:1DAE}, we show the  stress fluctuation correlations $C_{ijkl}(\theta)$ as a function of the angular variable $\theta \in [0,2\pi]$ for the same 2D experimental amorphous packing. To obtain the $\theta$ dependence of $C_{ijkl}$, we integrate $C_{ijkl}(q,\theta)$ over $q$, where $q$ is the magnitude of the Fourier space $q$-vector. The blue symbols indicate the experimental results averaged over $10$ independent datasets. The error bars indicate the corresponding statistical errors. In the same figure, the solid red lines show the theoretical predictions from 2D elasticity theory, Eq. \eqref{eq:isotropicCorrel}. The agreement between the experimental data and the theoretical predictions is evident. We notice that the theoretical predictions are based solely on elasticity theory. Despite this might seem surprising, we recall that, in the regime in which only quadrupole screening is relevant in amorphous solids, the mechanical response of athermal amorphous systems is well described by elasticity theory, \textit{albeit} with renormalized and parameter dependent elastic moduli. It is therefore not surprising that elasticity theory correctly captures the form of the correlations. On the other hand, whenever dipole screening becomes relevant, our formulae (and also the identical ones derived from the original VCT theory) will fail, since they will not be able to capture the emergent plastic screening length-scale that needs to be incorporate into the elasticity framework as done in \cite{PhysRevE.104.024904}. In the context of athermal granular matter, the regime in which dipole screening is negligible corresponds to the high-pressure limit (see, e.g., Fig.3 in \cite{PhysRevE.109.014902}).

    \begin{figure}[htbp]
    \centering
        \includegraphics[width=1.0\linewidth]{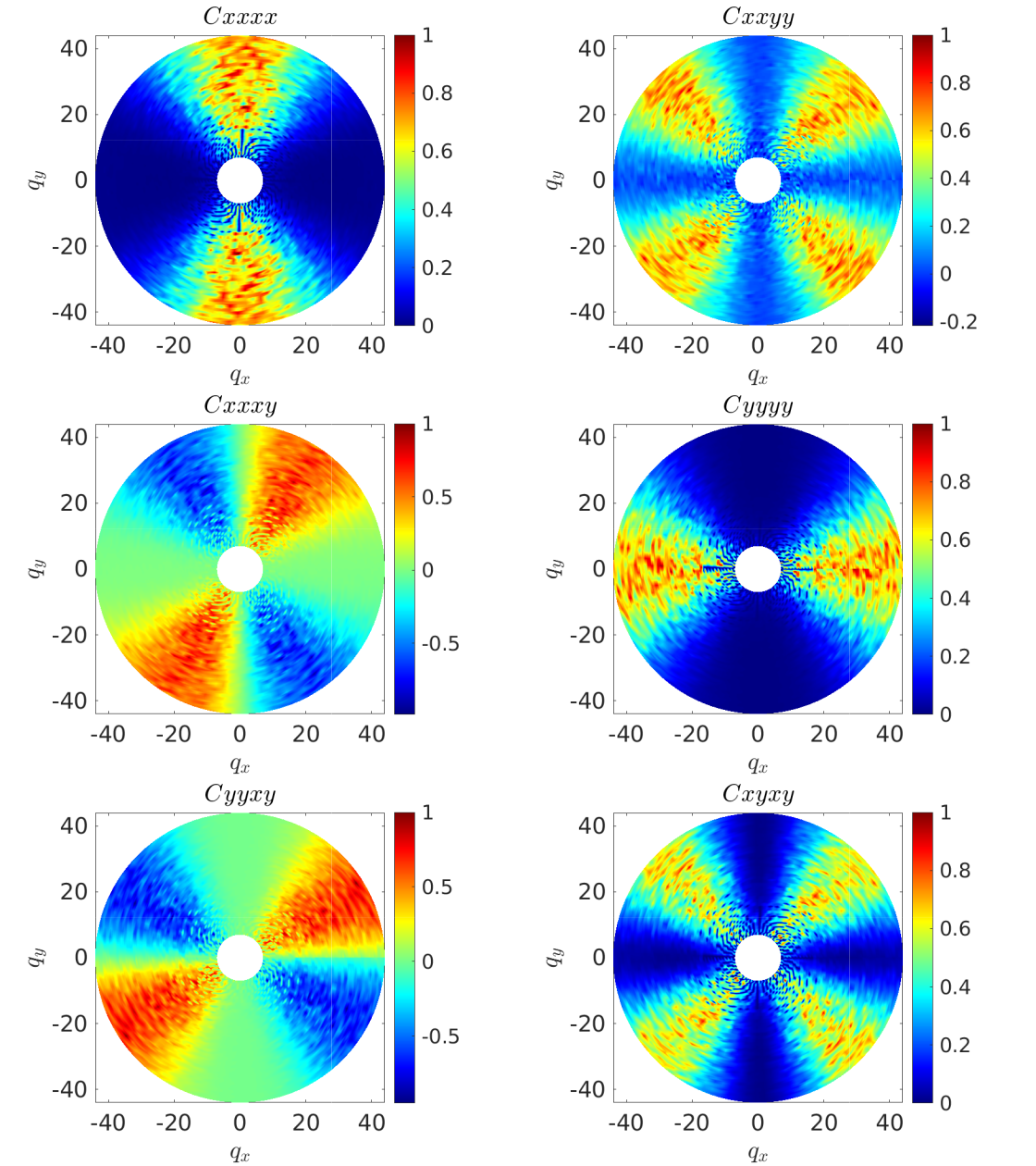}
    \caption{2D stress fluctuation correlations ($C_{xxxx}$, $C_{xxyy}$, $C_{xxxy}$, $C_{yyyy}$, $C_{yyxy}$, and $C_{xyxy}$) as a function of $\theta$, obtained from triangular lattice simulation (TLE in Fig. \ref{fig:gr}). The data are normalized by the maximum value of $C_{xxxx}$. Pinch-point singularities in the $q \rightarrow 0$ limit emerge in the correlation functions, since the latter are only a function of the angle $\theta$ in the continuous small $q$ regime.}
    \label{fig:2DTS}
\end{figure}

In order to confirm that the theoretical predictions in Eq. \eqref{eq:isotropicCorrel} hold also in isotropic crystalline systems, we consider a simulated 2D triangular lattice (see details in Section \ref{sim}). The crystalline structure is confirmed by the pair distribution function $g(r)$ shown in Fig. \ref{fig:gr} (TLS). The stress fluctuation correlations are shown in wave-vector space $(q_x,q_y)$ in Fig. \ref{fig:2DTS}. The expected angular dependence is already visible there. Moreover, one can observe the presence of pinch-point singularities, that were previously ascribed as a defining feature of amorphous solids. These singularities emerges as well in 2D isotropic solids and are in fact captured by Eq. \eqref{eq:isotropicCorrel}.\\

    \begin{figure}[htbp]
    \centering
    \includegraphics[width=1.0\linewidth]{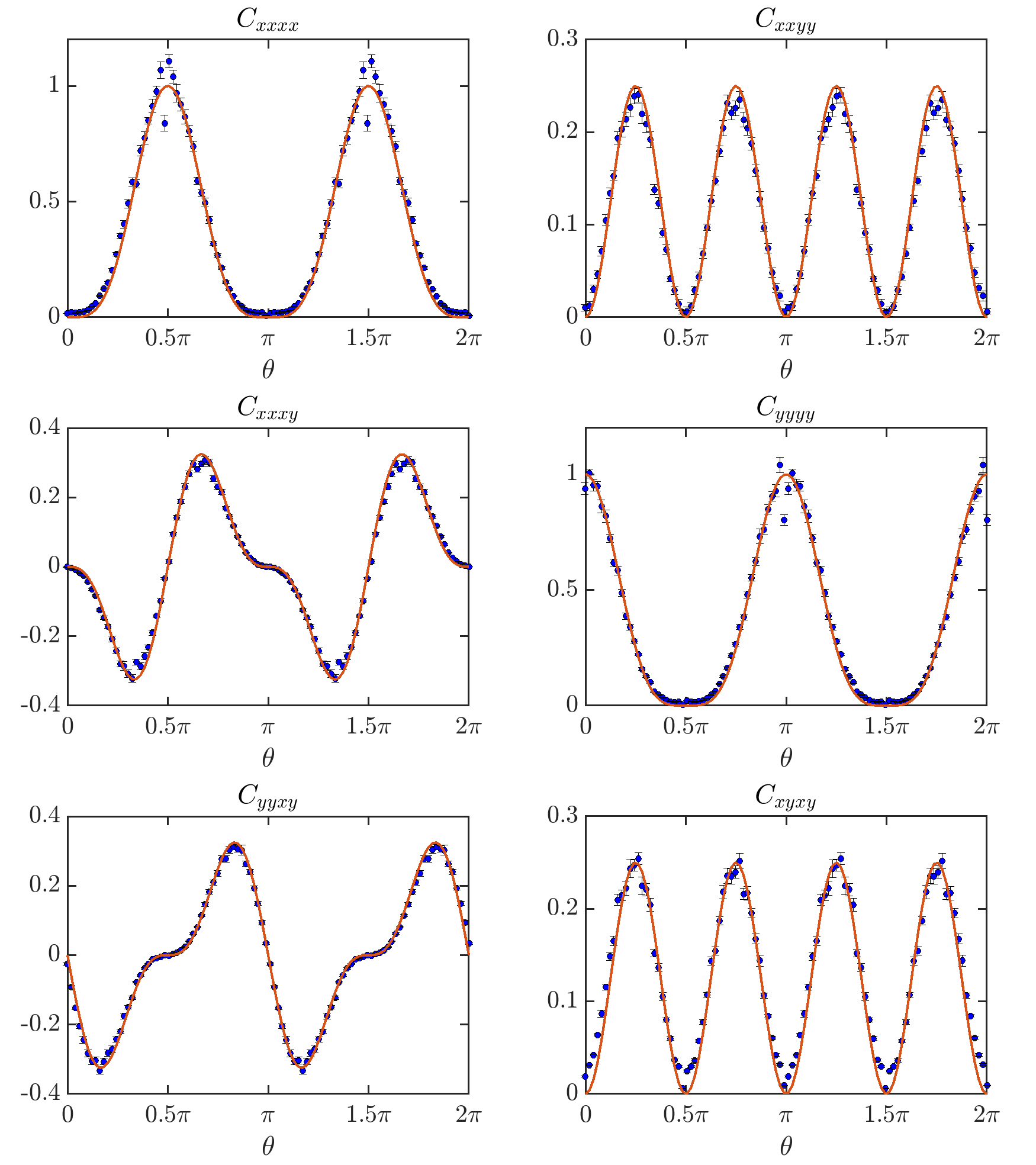}
    \caption{Normalized stress fluctuation correlations ($C_{xxxx}$, $C_{xxyy}$, $C_{xxxy}$, $C_{yyyy}$, $C_{yyxy}$, and $C_{xyxy}$) as a function of $\theta$, obtained from the triangular lattice simulation (TLS in Fig. \ref{fig:gr}). The normalization is performed by dividing each component by the maximum value of $C_{xxxx}$. The solid red lines represent the theoretical predictions in Eq. \eqref{eq:isotropicCorrel}. The error bars denote statistical error in 20 independent simulation samples.}
    \label{fig:1DTS}
\end{figure}
	
As a further test, we show the same correlations as a function of the angle $\theta$ in Fig. \ref{fig:1DTS}. The data are presented as averaged data points with error bars, calculated from 20 independent simulation groups. The blue symbols show the stress correlations extracted from the numerical simulations after averaging on $20$ independent simulation groups. The red solid lines are the theoretical predictions from elasticity theory, Eq. \eqref{eq:isotropicCorrel}. The agreement between the two is evident. We also notice that the form of the stress correlations for the isotropic 2D crystal shown in Fig. \ref{fig:1DTS} are identical to those obtained for the 2D amorphous experimental systems that are displayed in Fig. \ref{fig:1DAE}. This proves our initial hypothesis.

In Appendix \ref{app1}, we also show the results for a 2D experimental crystalline packing with triangular lattice. Unfortunately, due to the low statistics and experimental limitations, the quality of the experimental data is low and hence the comparison with the theoretical predictions is more difficult. Nevertheless, the same qualitative trend is observed.

   \begin{figure}[htbp]
    \centering
    \includegraphics[width=1.0\linewidth]{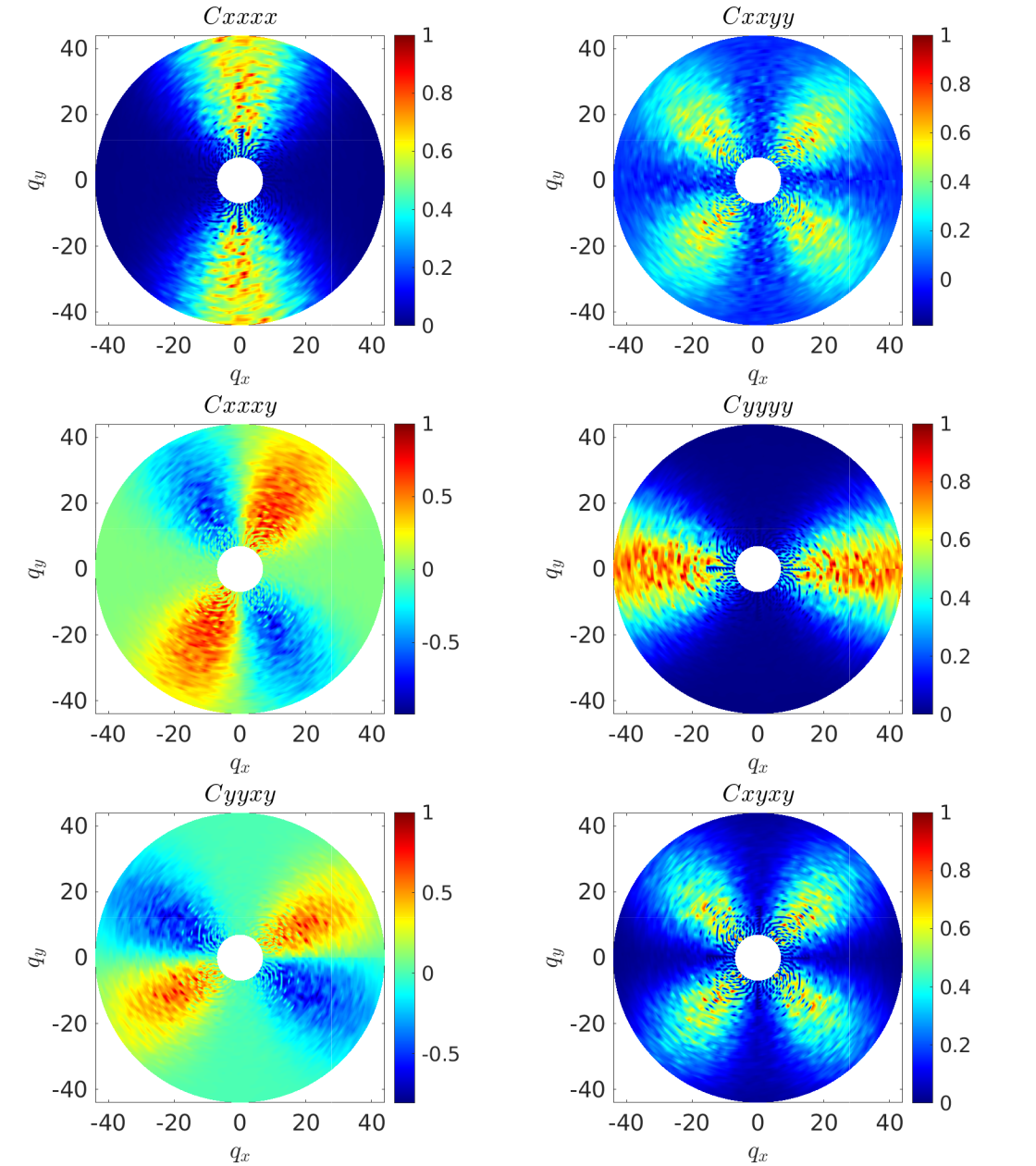}
    \caption{Normalized 2D correlation functions obtained from square lattice simulation. To improve statistics, $20$ groups of independent simulation are concerned. Pinch-point singularities in the $q \rightarrow 0$ limit are evident in all correlation functions.}
    \label{fig:2DSS}
\end{figure}

\begin{figure}[ht]
    \centering
    \includegraphics[width=\linewidth]{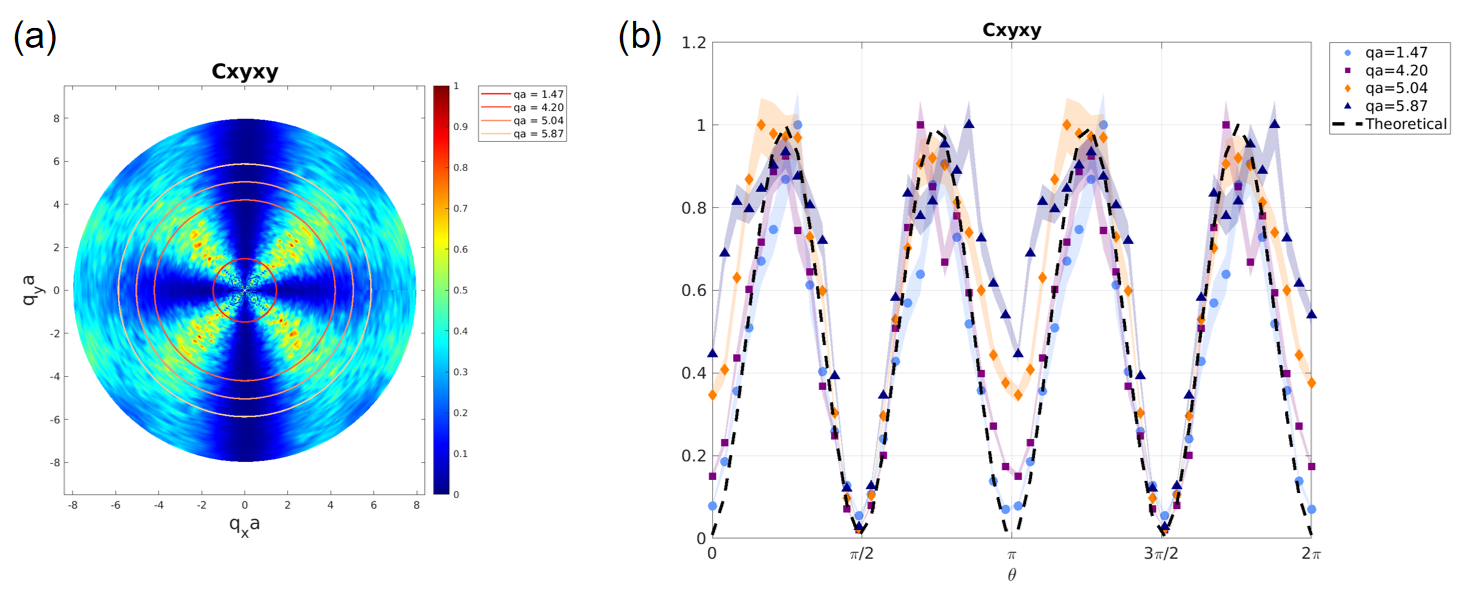}
    \caption{\textbf{(a)} 2D stress fluctuation correlation function $C_{xyxy}$ as a function of the normalized wave-vector $\textbf{q}a$ from the triangular lattice simulation. The various circles indicate the $q$-cuts shown in panel (b). \textbf{(b)} Angular dependence of the same correlation function at constant $q$. The dashed black line is the theoretical prediction from elasticity theory, Eq. \eqref{eq:isotropicCorrel} in the main text. The shaded regions are the statistical errors associated to the numerical data.}
    \label{fig:cuts}
\end{figure}

We emphasize that our theoretical framework is based on continuum elasticity theory, which is valid only in the long-wavelength limit, i.e., for wave vectors $q$ much smaller than the inverse lattice spacing $a$. To substantiate this point, Fig.\ref{fig:cuts} presents an analysis of the stress fluctuation correlation function $C_{xyxy}$ evaluated at different fixed values of $q$. In panel (a), we show $C_{xyxy}$ in the two-dimensional $(q_x,q_y)$ plane, with concentric circles indicating the $q$-cuts plotted in panel (b). As clearly seen in Fig.\ref{fig:cuts}(b), the theoretical prediction from Eq.~\eqref{eq:isotropicCorrel} matches the simulation results well at low $qa$. However, at higher $qa$, significant deviations appear, indicating the breakdown of the continuum approximation at shorter length scales.

\section{Stress-stress correlations in anisotropic crystals}
In order to prove the validity and generality of our theoretical predictions presented in Section \ref{thth}, we have also performed numerical simulations for a 2D crystalline system with square lattice. The crystalline structure is confirmed by the pair distribution function $g(r)$ shown in Fig. \ref{fig:gr} (SLS).

\begin{figure}[htbp]
    \centering
    \includegraphics[width=1.0\linewidth]{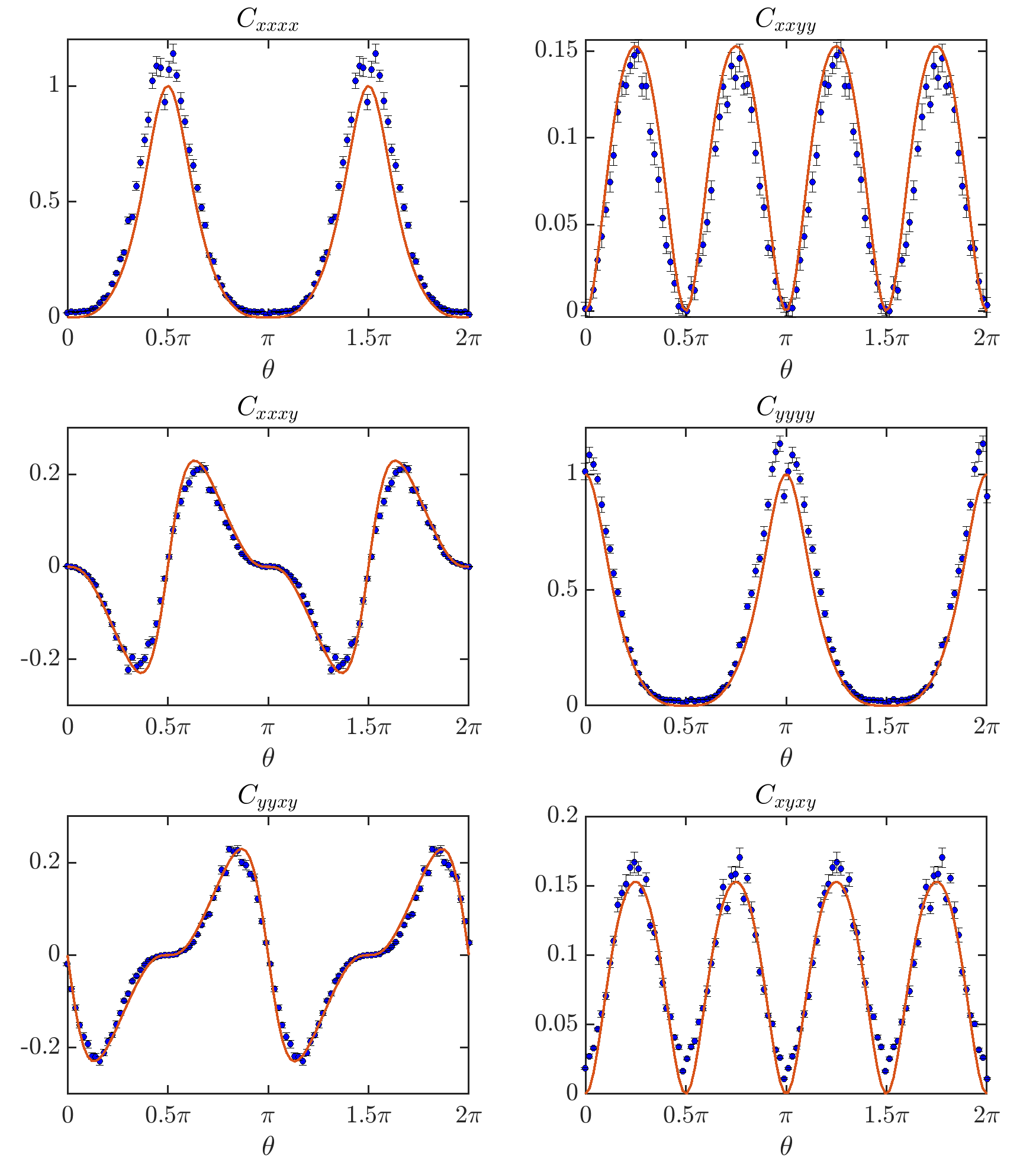}
    \caption{Normalized correlation functions obtained from square lattice simulation (SLS in Fig. \ref{fig:gr}). The solids lines correspond to the theoretical predictions, Eq. \eqref{eq:SquareCorrel}. The error bars denote statistical error in 20 independent simulation samples.}
    \label{fig:1DSS}
\end{figure}

In Fig. \ref{fig:1DSS}, we show the stress fluctuation correlation functions as a function of the angle $\theta$ and compare them with the theoretical predictions from elasticity theory, Eq. \eqref{eq:SquareCorrel}. The agreement is excellent.

    \section{Discussion}
    In this work, we revisited the problem of static stress-stress correlations in 2D solids with a particular emphasis on the contrast between amorphous and crystalline structures. After presenting a general derivation of these correlations from \textit{vanilla} elasticity theory \cite{landau2012theory,chaikin1995principles}, we have focus on two benchmark cases: (i) isotropic 2D solid (triangular crystalline structure) and (ii) anisotropic 2D solid (square lattice).

    By combining experimental and simulation data, we have demonstrated that the stress-stress correlations in 2D amorphous packing and in 2D isotropic crystals take the same form as a function of the angle $\theta$. Moreover, we have proved that this form can be directly derived using elasticity theory. As a direct consequence of these findings, we have shown that pinch-point singularities in stress-stress correlations are not an exclusive feature of amorphous solids as they generally appears also in crystals. Finally, we have explicitly shown that the same theoretical framework correctly predicts the same correlations in 2D anisotropic crystals, as expected.

    To avoid misunderstandings, we stress that the similarities between 2D amorphous and 2D isotropic crystalline systems are limited to the long-wavelength static stress correlators. Moreover, as already anticipated, they hold only in the regime in which amorphous systems can be described by a renormalized elastic framework, where only quadrupole screening is effective. Beyond the regime considered in this work, more sophisticated theories such as tensor gauge theories \cite{PhysRevE.106.065004} or anomalous elasticity \cite{PhysRevE.104.024904} have to be used. We also emphasize that the main difference between athermal amorphous solids and crystals lies in the ensemble average (Edwards vs Boltzmann) and leads to a different overall pre-factor in all the correlations, without affecting their overall functional form. 
    
    \section*{Acknowledgments}
    We thank Kabir Ramola, Michael Moshe and Bulbul Chakraborty for helpful discussions.
    JB and MB acknowledge the support of the Foreign Young Scholars Research Fund Project (Grant No.22Z033100604). MB acknowledges the sponsorship from the Yangyang Development Fund. YJW is financially supported by the National Natural Science Foundation of China (Grant No. 12472112). JS and JZ acknowledge the support of the NSFC (No.11974238 and No.12274291) and the Shanghai Municipal Education Commission Innovation Program under No. 2021-01-07-00-02-E00138. JS and JZ also acknowledge the support from the SJTU Student Innovation Center.

	\appendix
    
    \section{Isotropic 2D crystal: experiments}\label{app1}
\begin{figure}[htbp]
\includegraphics[width=1.0\linewidth]{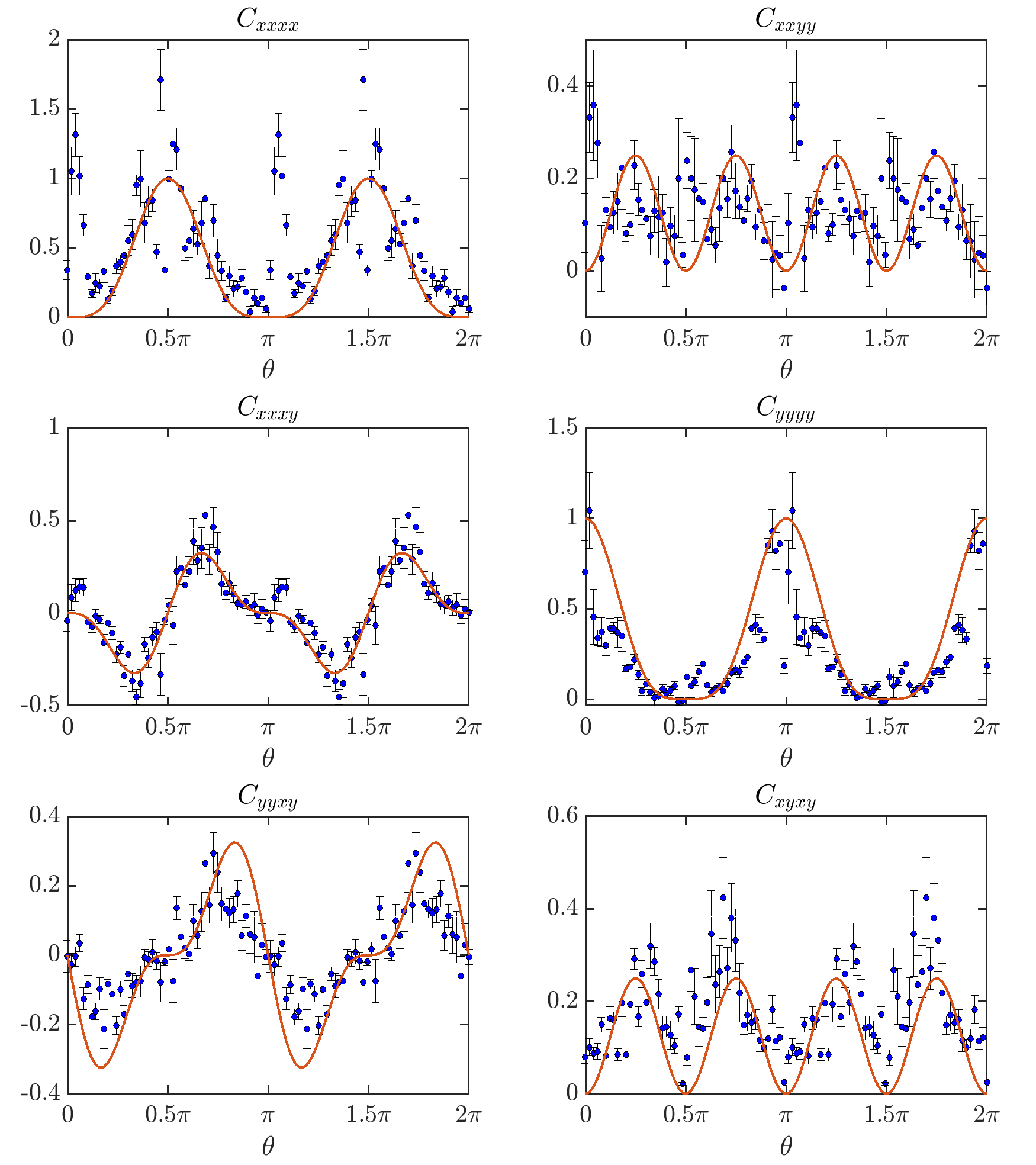}
\caption{Normalized stress fluctuation correlations $C_{ijkl}(\theta)$ obtained from the isotropic crystalline experimental packing (TLE in Fig. \ref{fig:gr}). The results are the ensemble average over $7$ independent experimental datasets and the error bars indicate the corresponding statistical errors. The solids lines are the theoretical predictions from Eq. \eqref{eq:isotropicCorrel}.}
    \label{fig:1DLE}
\end{figure}

    For completeness, in this Appendix we report also our experimental data on a 2D isotropic crystalline system with triangular lattice. The corresponding structure and pair distribution function are shown in Fig. \ref{fig:gr} in the main text (TLE). In Fig. \ref{fig:1DLE}, we show the stress correlations as a function of the angle $\theta$ and compare them to the theoretical predictions, Eq. \eqref{eq:isotropicCorrel}.
\end{document}